\documentclass[aps,physrev,reprint,superscriptaddress,longbibliography]{revtex4-2}

\usepackage{graphicx}
\usepackage{dcolumn}
\usepackage{bm}

\usepackage[colorlinks=true,allcolors=blue]{hyperref}
\usepackage{xurl}
\usepackage{amsmath}
\usepackage{mathtools}
\usepackage{amssymb}
\usepackage{bm}
\usepackage[normalem]{ulem}
\usepackage{tikz}

\usepackage{microtype}
\usepackage{subcaption}
\usepackage{soul}
\usepackage{bbm}
\usepackage{bbold}
\usepackage{multirow}
\usepackage{array}
\usepackage{stackengine}
\usepackage{braket}
\usepackage{comment}
\usepackage{appendix}

\begin{document}

\title{Contextuality and Chaos}

\author{Sanchit Srivastava}
\email{sanchit.srivastava@uwaterloo.ca}
\affiliation{Institute for Quantum Computing, University of Waterloo, Waterloo, Ontario, Canada N2L 3G1}
\affiliation{Department of Physics and Astronomy, University of Waterloo, Waterloo, Ontario, Canada N2L 3G1}
\author{Shohini Ghose}
\affiliation{Institute for Quantum Computing, University of Waterloo, Waterloo, Ontario, Canada N2L 3G1}
\affiliation{Department of Physics and Astronomy, University of Waterloo, Waterloo, Ontario, Canada N2L 3G1}
\affiliation{Department of Physics and Computer Science, Wilfrid Laurier University, Waterloo, Ontario, Canada N2L 3C5}

\begin{abstract}

Classical chaos is marked by an extreme sensitivity to initial conditions, where infinitesimally close trajectories separate exponentially over time. In quantum mechanics, however, unitary evolution and the uncertainty principle preclude such behavior, necessitating alternative approaches to identifying chaos in quantum systems. One must therefore seek quantum features that can indicate the emergence of chaos in the classical limit. Here, we show that contextuality, a quantum property that defies classical explanations,  can serve as a signature of chaos. For a spin system undergoing chaotic dynamics, we demonstrate that violations of Bell-type inequality can effectively differentiate regular and chaotic regions of the phase space, suggesting that the nonclassicality of the system underpins signatures of chaos. 

\end{abstract}

\maketitle

\section{Introduction} 
Chaotic dynamics is a feature of many classical systems, characterized by the exponential divergence of nearby trajectories in phase space. States of a classical system are specified by dynamical variables, and can be represented as points on a phase space. The evolution of the system in time is then described by trajectories in this phase space. In chaotic systems, neighboring trajectories diverge exponentially, leading to a sensitive dependence on initial conditions \cite{ottChaosDynamicalSystems2002}. This sensitivity is quantified by the Lyapunov exponent \cite{lyapunovGeneralProblemStability1994}, which measures the rate of divergence of nearby trajectories. Such a description of the dynamics in terms of trajectories is incompatible with quantum theory, wherein the notion of trajectories is not well defined due to the uncertainty principle. Moreover, non-dissipative quantum systems evolve unitarily, leading to a linear evolution of the state. This linearity implies that the exponential divergence of trajectories, a hallmark of chaos, is absent in quantum systems. For quantum systems showing chaotic dynamics in the classical limit, a characterization of chaos must be sought in terms of features intrinsic to the quantum system. 

Since chaos is a property of the dynamics of the system, it is natural to look for dynamical signatures such as quantum correlations and entropy. To this end, various signatures have been proposed, such as entanglement \cite{wangEntanglementSignatureQuantum2004,ghoseEntanglementDynamicsChaotic2004} and quantum coherence \cite{anandQuantumCoherenceSignature2021a}. Since we are interested in features inherent to the quantum system, a natural question arises: can a quantification of how nonclassical a quantum system is be a signature of chaos? One way to study the noclassicality of a quantum system is contextuality, a feature of quantum systems that allows them to generate measurement statistics which cannot be reproduced by any classical model \cite{budroniKochenSpeckerContextuality2022}. In this work, we show that contextuality, as measured via violations of a Bell-type inequality \cite{klyachkoSimpleTestHidden2008}, captures the structure of the phase space for a paradigmatic chaotic system, the kicked top \cite{haakeClassicalQuantumChaos1987a}. Our results suggest that it is the nonclassicality of the quantum system which acts as a signature of chaos, and hence provides a general explanation of some previously identified signatures of chaos.

This paper is structured as follows. In Section II, we discuss the kicked top model and describe its classical and quantum evolution. We then provide a brief overview of quantum contextuality and introduce the KCBS inequality as a test for contextuality in spin-1 systems \cite{klyachkoSimpleTestHidden2008}. In Section III, we show how normalized violations of this inequality can be used to identify regular and chaotic regions of the phase space of the kicked top. Finally, in Section IV we comment on how nonclassicality as a signature of chaos can provide a unified framework for understanding some previously proposed signatures of chaos like entanglement, and discuss further implications of our results.

\section{background}
\subsection{The kicked top model}\label{sec:kicked_top}
The kicked top is a finite-dimensional spin system periodically driven that exhibits chaotic dynamics in the classical limit. The evolution of the system is governed by the time-dependent Hamiltonian \cite{haakeClassicalQuantumChaos1987a}

\begin{equation}\label{eq:qkt}
    H = \hbar \frac{p J_y}{\tau} + \hbar \frac{\kappa J_z^2}{2j} \sum_{n=-\infty}^{\infty} \delta(t-n\tau),
\end{equation}
where $J_x, J_y, J_z$ are the angular momentum operators: $[J_p,J_q] = i \epsilon_{pqr}J_r$ and $j$ represents the size of the spin system. The first term in the Hamiltonian describes a precession of the spin about the $y$ axis by an angle $p$ in time period $\tau$. The second term is a periodic kick along the $z$ axis, which is a delta function pulse of strength $\kappa$ acting at times $t = n\tau$. In the limit of $j\rightarrow \infty$, the classical equations of motion for the kicked top can be obtained using Hamilton's equations. In terms of the dynamical variables $X = J_x/j, Y = J_y/j$ and $Z = J_z/j$, the classical evolution for the choice of $\tau=1$ and $p = \pi/2$ is given by the stroboscopic map

\begin{align}
    X_{n+1} &= Z_n \cos(\kappa X_n \tau) - Y_n \sin(\kappa X_n \tau) \nonumber \\
    Y_{n+1} &= Y_n \sin(\kappa X_n \tau) + Z_n \cos(\kappa X_n \tau) \nonumber \\
    Z_{n+1} &= -X_n
\end{align} 
where $n$ is the number of elapsed time periods. The kicking strength $\kappa$ acts as a chaoticity parameter for the system. As $\kappa$ is varied, the classical phase space shows a transition from regular dynamics ($\kappa \leq 2.1$) to a mixture of regular and chaotic behavior ($2.1 \leq \kappa \leq 4.4$) to fully chaotic dynamics for ($\kappa \geq 4.4$) \cite{kumariQuantumclassicalCorrespondenceVicinity2018}. 
Since the total angular momentum of the system remains conserved under this Hamiltonian, the dynamical variables are constrained by $X^2 + Y^2 + Z^2 = 1$. Hence, the phase space points of the system can be parameterized in terms of their polar coordinates $(\theta,\phi)$. Fig.\ref{fig:contour}a and Fig.\ref{fig:contour}c show the classical phase space for $\kappa = 0.5$ and $\kappa = 2.5$ respectively. 

In the quantum regime, the dynamics of the quantum kicked top (QKT) over one time period can be described by the Floquet unitary operator

\begin{equation}
    U  = \exp\left(-i\frac{\kappa}{2j}J_z^2\right) \exp\left(-i\frac{p}{\tau}J_y\right).
\end{equation}

Quantum dynamics of the kicked top are obtained by the evolution of spin coherent states under this Floquet unitary. Spin coherent states are minimum uncertainty states of the spin system which saturate the Heisenberg uncertainty relations between the angular momentum operators \cite{radcliffePropertiesCoherentSpin1971}. A spin coherent state centred at a phase space point $(\theta,\phi)$ can be prepared by rotating the maximum spin state along the $z$-axis by an appropriate rotation operator $R(\theta, \phi)$. Spin coherent states are the quantum analogues of the classical angular momentum states \textemdash they approximate points on the phase space in the classical limit of $j\rightarrow \infty$.  

\subsection{Contextuality} \label{sec:contextuality} 
Outcomes of quantum measurements are probabilistic in nature. This means that different runs of the same measurement, performed on identically prepared quantum systems, can yield different outcomes.  
Hidden variable models are attempts at explaining this apparent randomness by assuming that measurement outcomes are predetermined by some underlying classical variables with a probability distribution over them \cite{bellEinsteinPodolskyRosen1964}. The Kochen-Specker theorem and the two theorems of John Bell are $no-go$ hidden variable theorems\cite{merminHiddenVariablesTwo1993}. They establish that, under certain assumptions, it is impossible to explain the predictions of quantum mechanics using hidden variables. 

Quantum contextuality is a feature of quantum theory whereby measurements of quantum observables cannot simply be thought of as revealing pre-existing properties of the system. Any attempt to do so with a realistic hidden-variable theory leads to value assignments that are dependent on the choice of other compatible measurements being performed (i.e, the measurement $contexts$). 
For measurement scenarios where an observable appears in more than one context, a classical model would need to be noncontextual - the value assigned to the observable would be independent of the context in which it is being measured. However, in quantum mechanics, we can get probability distributions over the joint measurement outcomes of these contexts which are incompatible with any such classical value assignments. These distributions can only be achieved if the assignments are context dependent. 

Early demonstrations of contextuality were primarily logical proofs, based on identifying specific sets of vectors where any noncontextual value assignment resulted in a logical contradiction \cite{budroniKochenSpeckerContextuality2022,merminHiddenVariablesTwo1993}. Later, statistical approaches to proving contextuality emerged, analogous to statistical proofs of nonlocality. These methods involve identifying quantum states that produce measurement statistics incompatible with classical models. One such state-dependent proof is the Bell-type inequality formulated by Klyachko, Can, Binicioglu, and Shumovsky (KCBS), which serves as a test for the existence of hidden variable models for the measurement statistics of a spin-1 system \cite{klyachkoSimpleTestHidden2008}.

The KCBS inequality can be formulated in the following way. 
Consider a set of five unit vectors $r_i \in \mathbb{R}^3$ such that $r_i \perp r_{i+1} \forall i \in \mathbb{Z}_4$.  
Corresponding to each vector $r_i$, define spin operators $S_{i}$ which generate rotations of the spin states about the $r_i$ axis. These operators have eigenvalues 0,1 and -1.  
Let $\Pi^0_i$ be the projectors onto the zero-eigenspace of $S_{r_i}$. Note that $[\Pi^0_i,\Pi^0_{i+1}] = 0$ $\forall i\in \mathbb{Z}_4$ 
The projectors $\{\Pi^0_i\}$ and their orthogonality relations can be represented by a pentagon graph, where the five vertices are the projectors and the edges are the orthogonality relations. 
Since the eigen values of these projectors are either $0$ or $1$, any noncontextual value assignment on these projectors can be thought of as a coloring of the vertices of the graph with two colors. If we represent the $1$ eigenvalue as green and the $0$ eigenvalue as red, then the orthogonality relations imply that no two adjacent vertices can be colored green. 
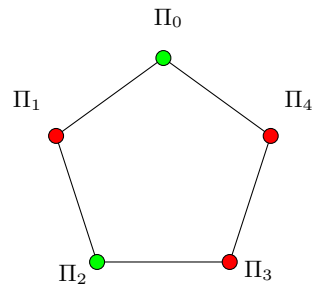
\begin{figure}[h!]
    \centering
    \label{fig:pentagon}
    \begin{tikzpicture}[scale=1]
        \foreach \i in {0,...,4} {
            \pgfmathsetmacro{\angle}{90 + 72*\i}
            \coordinate (P\i) at (\angle:1.5cm);
        }
        \draw (P0) -- (P4) -- (P3) -- (P2) -- (P1) -- cycle;
        \node[draw, circle, fill=green, inner sep=2pt, label={[shift=(70:2mm)]$\Pi_0$}] at (P0) {};
        \node[draw, circle, fill=red, inner sep=2pt, label={[shift=(162:4mm)]$\Pi_1$}] at (P1) {};
        \node[draw, circle, fill=green, inner sep=2pt, label={[shift=(238:6mm)]$\Pi_2$}] at (P2) {};
        \node[draw, circle, fill=red, inner sep=2pt, label={[shift=(310:6mm)]$\Pi_3$}] at (P3) {};
        \node[draw, circle, fill=red, inner sep=2pt, label={[shift=(18:4mm)]$\Pi_4$}] at (P4) {};
    \end{tikzpicture}
    \caption{A possible coloring of the graph representing the orthogonality relations of the projectors $\{\Pi^0_i\}$.} 
\end{figure}

Due to the above constraint, at most two vertices maybe colored green in any coloring of the graph. This leads to the inequality

\begin{equation}\label{eq:KCBS}
    \langle\beta_{QM}\rangle:= \sum_{i \in \mathbb{Z}_4} \langle\Pi^0_i\rangle \leq 2.
\end{equation}

We will refer to $\beta_{QM}$ as the KCBS operator. The above inequality can be reformulated in terms of the expectation values of the spin operators $S_{r_i}$, to obtain the original form of the KCBS inequality: 

\begin{equation}\label{eq:KCBS2}
    \sum_{i \in \mathbb{Z}_4} \langle S^2_{r_i}\rangle \geq 3.
\end{equation}

Any noncontextual model of the measurement statistics of a spin-$1$ system must satisfy the above inequality. It was shown that the maximum quantum expectation value of the KCBS operator,  $\langle \beta_{QM}\rangle_{max}$ can reach up to $\sqrt{5}$ \cite{klyachkoSimpleTestHidden2008}, which violates the inequality. The KCBS inequality is a part of a more general family of inequalities called the CSW inequalities, which are graph-theoretic inequalities that can detect contextuality \cite{cabelloNonContextualityPhysicalTheories2010a}. Like in the case of their nonlocality counterparts, the extent of violations of these inequalities can be used to quantify the amount of contextuality in a quantum system \cite{desilvaGraphtheoreticStrengthsContextuality2017}.

\section{Contextuality as a signature of chaos}
The maximum violations of the KCBS inequality occur for the states $\ket{0}_r$, which are the zero eigenstates of the spin operators $S_{r}, r \in \mathbb{R}^3$. By contrast, the states $\ket{1}_r$ never violate the inequality. The states $\ket{1}_r$ are exactly the spin coherent states mentioned in Section \ref{sec:kicked_top} . This means that the initial states of the QKT system, before the action of the Floquet operator, are classical in the sense of not exhibiting contextuality. As the states evolve under the Floquet operator, they deviate from the classical initial states and can start exhibiting contextuality. 

The classical dynamics of the kicked top feature a variety of special points in the phase space, such as fixed points and periodic orbits \cite{haakeClassicalQuantumChaos1987a}. Of special interest are the fixed points $FP_1 = (\pi/2,\pi/2)$, $FP_2 = (-\pi/2, \pi/2)$ and the period-$4$ orbit $P_4 = (\pi/2,0) \rightarrow (\pi,0) \rightarrow (\pi/2,\pi) \rightarrow (0,0) \rightarrow (\pi/2,0)$. $FP_1$, $FP_2$ and $P_4$ persist at all values of $\kappa$, undergoing bifurcations as $\kappa$ is varied, which alter the stability of these points \cite{kumariQuantumclassicalCorrespondenceVicinity2018}. The behavior of the signature at these points must be analyzed separately. Furthermore, the form of the Floquet unitary gives rise to periodicities in the QKT evolution with respect to $\kappa$ which are absent in the classical dynamics \cite{bhosalePeriodicityQuantumCorrelations2018}. To account for these periodicities, we restrict our analysis to $\kappa$ values in the range $[0,\pi]$.

In the classical dynamics of the system, the nature of a phase space point is characterized by its trajectory as it evolves over time. In the quantum regime, we associate to these trajectories the quantity
\begin{equation*}
    K(\theta,\phi) = \lim_{N\rightarrow \infty} \frac{1}{N} \sum_{n=0}^{N} \beta'(\rho(\theta,\phi,n)),
\end{equation*}

where $\rho(\theta, \phi, n) = U^n \ket{\theta,\phi} \bra{\theta,\phi} U^{\dagger n}$ is the state of the system initialized in a spin coherent state centred at $(\theta,\phi)$ after $n$ steps of the Floquet evolution, and $\beta'(\rho)$ is the normalized violation of the KCBS inequality for state $\rho$ defined as 

\begin{equation}
    \beta'(\rho) =
    \begin{cases}
    \frac{\langle \beta_{QM}\rangle_{max} - 2}{\sqrt{5}-2}, &  \langle \beta_{QM}\rangle_{max} > 2, \\
    0, &  \langle \beta_{QM}\rangle_{max} \leq 2.
    \end{cases} 
\end{equation}

The maximum violations are normalized by the factor of $\sqrt{5}-2$ to ensure that the quantity $K(\rho)$ lies in the range $[0,1]$. Hence, $K(\theta, \phi)$ is the average normalized KCBS violation of a state's trajectory under QKT evolution. We can leverage the temporal periodicity of the QKT evolution and compute the quantity $K(\theta,\phi)$ over a finite range of $N$ (depending on the range of $\kappa$ under consideration) which is sufficient to capture the entire quantum dynamics of the system \cite{anandQuantumRecurrencesKicked2024a} . 

\begin{figure}[h]
    \centering
    
    \begin{subfigure}{0.25\textwidth}
        \includegraphics[width=\textwidth]{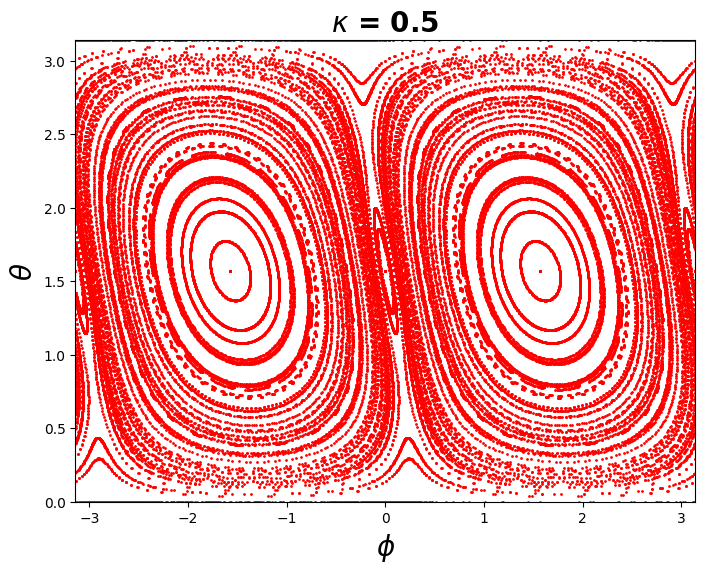}
        \caption{Classical phase space for $\kappa = 0.5$}
    \end{subfigure}
    \begin{subfigure}{0.25\textwidth}
        \includegraphics[width=\textwidth]{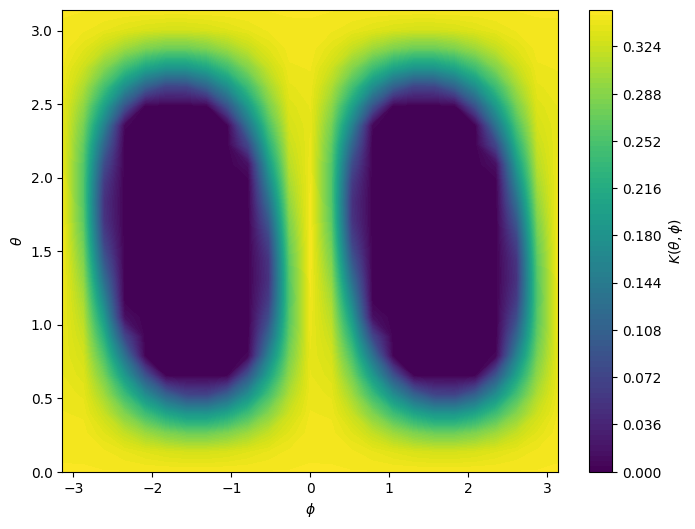}
           \caption{Contour plot of $K(\theta,\phi)$ for $\kappa = 0.5$} 
    \end{subfigure}
    
    \vspace{0.5cm}
    
    \begin{subfigure}{0.25\textwidth}
        \includegraphics[width=\textwidth]{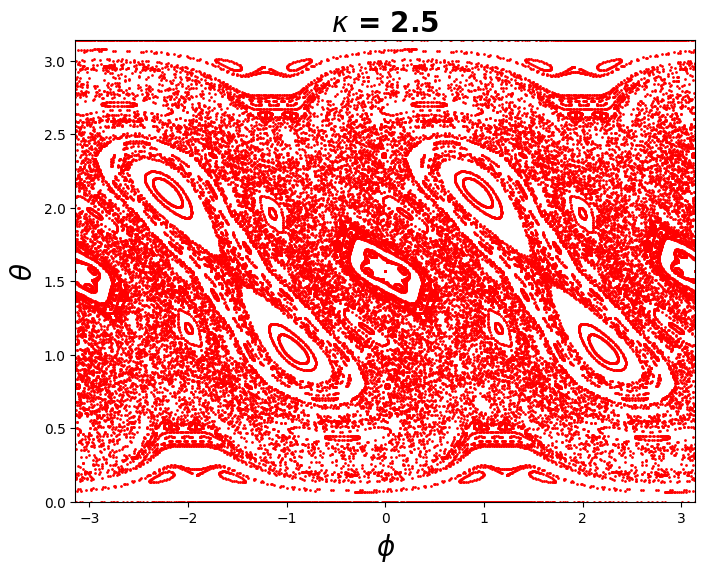}
        \caption{Classical phase space for $\kappa = 2.5$}
    \end{subfigure}
    \begin{subfigure}{0.25\textwidth}
        \includegraphics[width=\textwidth]{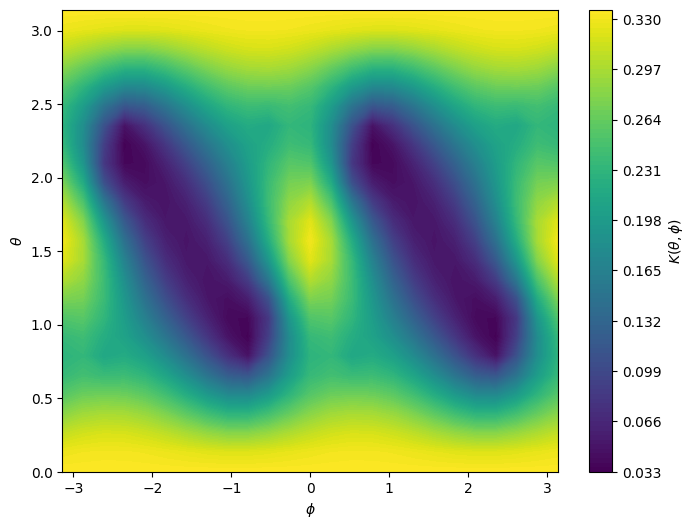}
        \caption{Contour plot of $K(\theta,\phi)$ for $\kappa = 2.5$}
    \end{subfigure}
    
    \caption{Comparison of classical and quantum dynamics.}\label{fig:contour}
\end{figure}

For a state $\rho$, the maximum expectation value of the KCBS operator $\langle\beta_{QM}\rangle_{max}$ can be computed numerically using the \texttt{scipy.optimize} module in Python \cite{virtanenSciPy10Fundamental2020}. Here the maximization has been performed over the spin operators that define the projectors of $\beta_{QM}$. Starting with 625 different spin coherent states spread over the sphere, we calculate $\beta'$ for each initial point $(\theta,\phi)$ after each time step $n$ and plot the average $K(\theta,\phi)$ over $50$ steps as a contour over the $\theta-\phi$ plane. In Fig.\ref{fig:contour}, we compare the classical phase space (left column) with the contour plots of $K(\theta,\phi)$ (right column) for two different values of $\kappa$ : $\kappa = 0.5$, where the dynamics are regular, and $\kappa = 2.5$, where the phase space shows both regular and chaotic behavior. 

We see that the contour plots of $K(\theta,\phi)$ (Fig.\ref{fig:contour}b and \ref{fig:contour}d) capture the structure of the phase space for the classical kicked top. The islands of regular dynamics in the phase space show systematically lower values than the chaotic regions.
Further, we note that the maximum violations averaged over the phase space increases with the chaoticity parameter $\kappa$. 
\begin{figure}[h!]
    \centering
    \includegraphics[width=0.45\textwidth]{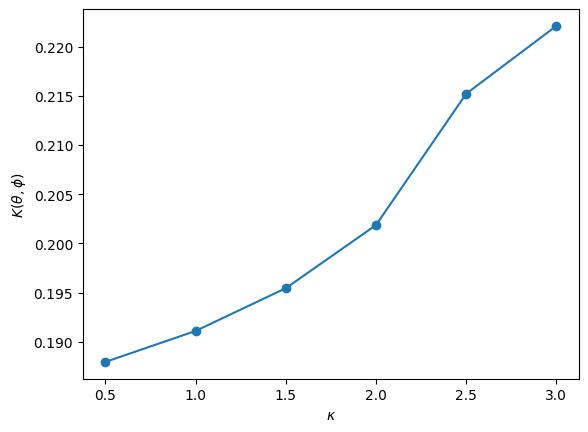}
    \caption{$K(\theta,\phi)$ averaged over the 625 initial points and 50 kicks as a function of $\kappa$.}\label{fig:kvskappa}
\end{figure}
Fig.\ref{fig:kvskappa} shows the average of $K(\theta,\phi)$ over the 625 initial points oon the phase space as a function of $\kappa$. We see that the average violation increases monotonically with $\kappa$. These observations show that the averaged KCBS violation can be used as a signature of chaos in the QKT. 

We now turn our attention to the special cases of the aforementioned fixed points and periodic orbits in the classical evolution of the kicked top. An analysis of the evolution of Husimi distributions for the spin coherent states at points $FP_1$ and $FP_2$ indicates that the quantum evolution of these points closely resembles the classical dynamics when $\kappa < 0$. This means that under the Floquet evolution, these states deviate very little from the original spin coherent states, and hence we see that the KCBS violations at these points are close to zero. By contrast, for $j =1$ the quantum dynamics of the states lying on the period-4 orbit $P_4$ show a significant deviation from the classical dynamics \cite{kumariQuantumclassicalCorrespondenceVicinity2018,anandSimulatingQuantumChaos2024}. This is reflected in the KCBS violations, where we see much higher violations for these points despite them constituting a regular orbit in the classical phase space.  

\section{Discussion}


In this work, we have explored the relationship between contextuality, as measured by violations of the KCBS inequality, and chaotic behavior in the quantum kicked top (QKT). Our numerical analysis demonstrates that the time-average, normalized KCBS violations serve as an effective means of distinguishing regular and chaotic regions in the system's phase space, and are hence a useful signature of chaos. 

Over the years, numerous quantum signatures of chaos have been proposed, each with varying degrees of applicability across different models \cite{anandQuantumCoherenceSignature2021a,passarelliChaosMagicDissipative2024,schackHypersensitivityPerturbationQuantum1994,angeloRecoherenceEntanglementDynamics2001,furuyaQuantumDynamicalManifestation1998}. Of particular interest in the case of spin systems is entanglement. The Hilbert space of a spin-$j$ system is isomorphic to the permutationally invariant subspace of an ensemble of $2j$ qubits \cite{harrowChurchSymmetricSubspace2013}. For the quantum kicked top, entanglement between the qubits in this representation of the quantum states has been shown to be an effective signature of chaos \cite{ghoseEntanglementDynamicsChaotic2004,wangEntanglementSignatureQuantum2004}. The nonlinear operator $J_z^2$ in the Hamiltonian transforms as a two qubit operator $J_z^2 = \sum_{i,j} \sigma^i_z \otimes \sigma^j_z$ in the qubit picture, where $\sigma^i_z$ denotes the  spin-$z$ operator on the $i$th qubit. Since this nonlinearity drives the onset of chaos, it is natural to expect that entanglement between the qubits would be a signature of chaos. However, considering the kicked top as just a spin particle undergoing rotations and twists, it is desirable to seek an explanation for this entanglement signature in terms of physically measurable properties of this particle, like the expectation values of spin operators.

For spin-1 systems, the states of the form $\ket{0}_r$ which maximally violate the KCBS inequality, correspond to the maximally entangled states in the qubit picture. By contrast, the spin coherent states, which never violate the inequality, correspond to separable states. The authors of \cite{klyachkoSimpleTestHidden2008} state that these observations suggest that the nonclassicality of spin-1 states captured by the KCBS inequality may be a consequence of entanglement between the internal degrees of freedom of the spin particle. This perspective reframes the role of entanglement as a signature of chaos, not merely as a correlation measure in some representation, but as a manifestation of the intrinsic nonclassicality of the quantum system.  

In the quantum kicked top, when $\kappa = 0$, the Floquet unitary simplifies to a rotation about the $y-$axis, causing initial spin coherent states to evolve into other spin coherent states, which remain noncontextual. This behavior is analogous to the action of a stabilizer operation on stabilizer states. For $\kappa > 0$, the nonlinear term introduces a twist operation of the form $\exp(-i\frac{\kappa}{2j}J_z^2)$ , which deviates from stabilizer dynamics. This shift is akin to the action of a non-stabilizer operation, suggesting a deeper link between emergence of chaos and the classical intractability of quantum dynamics, which has been identified as a signature of chaos in some recent works \cite{passarelliChaosMagicDissipative2024,gotoChaosMagic2021,leoneQuantumChaosQuantum2021}. In \cite{anandQuantumCoherenceSignature2021a} quantum coherence was shown to be a signature of chaos. Quantum coherence is a measure of the ability of a state to exist in a superposition of some preferred basis states, and can be considered as an indicator of the nonclassicality of the state\cite{wagnerInequalitiesWitnessingCoherence2024a}. 

These observations reinforce the perspective that it may indeed be the nonclassicality of the quantum system which acts as a signature of chaos. While we have provided numerical evidence for a specific model, a more general study of the relationship between chaotic dynamics and contextuality in quantum systems is warranted. Such an analysis could provide a unified framework for understanding signatures of chaos in quantum systems, and may offer new insights into unpredictable behavior of chaotic systems.
\section*{Acknowledgments}
This work was supported by the  Natural Sciences and Engineering Research Council of Canada. Wilfrid Laurier University is located in the traditional territory of the Neutral, Anishnawbe and Haudenosaunee peoples. We thank them for allowing us to conduct this research on their land.  

\bibliography{main}
\appendix
\end{document}